\documentclass[DIV11]{scrartcl}

\usepackage[utf8]{inputenc}

 \usepackage{musixtex}
\usepackage{natbib} 

\usepackage{etex} 
 \usepackage{amsmath}
 \usepackage{graphicx}
 \usepackage{fancybox}
\usepackage{pgfplots}
\usepackage{url}
\pgfplotsset{compat=1.3} 
 \usepackage{subfigure}
 \usepackage[draft,inline,marginclue]{fixme}
 \usepackage{booktabs} 
\usepackage{float}

\usepackage{setspace}

\usepackage{authblk}
\usepackage{blindtext}

\author[,$\dagger$,1,2]{Kat R. Agres\thanks{These authors contributed equally to this work and share first authorship.}}
\author[$\textasteriskcentered$,1]{Adyasha Dash}
\author[3]{Phoebe Chua}\small
\affil[1]{\small Yong Siew Toh Conservatory of Music, National University of Singapore, Singapore}
\affil[2]{\small Centre for Music and Health, National University of Singapore, Singapore}
\affil[3]{Augmented Human Lab, DISA, National University of Singapore, Singapore}
\affil[$\dagger$ ]{Corresponding author: Kat R. Agres, katagres@nus.edu.sg}

\date{}

\title{AffectMachine-Classical: A novel system for generating affective classical music} 

\begin{document}



\maketitle

\begin{abstract}

This work introduces a new music generation system, called AffectMachine-Classical, that is capable of generating affective Classic music in real-time. AffectMachine was designed to be incorporated into biofeedback systems (such as brain-computer-interfaces) to help users become aware of, and ultimately mediate, their own dynamic affective states. That is, this system was developed for music-based MedTech to support real-time emotion self-regulation in users. We provide an overview of the rule-based, probabilistic system architecture, describing the main aspects of the system and how they are novel. We then present the results of a listener study that was conducted to validate the ability of the system to reliably convey target emotions to listeners. The findings indicate that AffectMachine-Classical is very effective in communicating various levels of Arousal ($R^2 = .96$) to listeners, and is also quite convincing in terms of Valence ($R^2 = .90$). Future work will embed AffectMachine-Classical into biofeedback systems, to leverage the efficacy of the affective music for emotional well-being in listeners.


\textbf{Keywords:} automatic music generation system, algorithmic composition, music MedTech, emotion regulation, listener validation study 
\end{abstract}

\section{Introduction}

There is now overwhelming evidence that music supports health and well-being in various ways, from motivating physical activity, to promoting mental health and fostering social connection \citep{fancourt2019evidence, macdonald2013music}.
Music is particularly effective for supporting and mediating emotion states. Indeed, one of the primary reasons people report listening to music is to change or enhance their emotions \citep{thayer1994self, lonsdale2011we, saarikallio2012development}. Given the affordances of music to support health and well-being, as well as advances in machine learning and computational techniques, there has recently been a call to action to compose music with the use of computational technologies for healthcare applications \citep{agres2021music}. 
These types of generative music composition systems show promise in delivering cost-effective, non-invasive, non-pharmaceutical methods for helping individuals improve their emotion states. Given the global mental health crisis (e.g, nearly 20\% of adults in the USA live with a mental illness, or 52.9 million Americans in 2020\footnote{Statistics from the National Institute of Mental Health: https://www.nimh.nih.gov/health/statistics/mental-illness}), music medtech systems are projected to be extremely valuable tools for supporting emotional wellness, and mental health more broadly.


More generally, in an age where computational systems are now being used extensively to generate impressive natural language and visual art, such as the technologies available through OpenAI\footnote{https://openai.com/}, it is no surprise that there has been a recent surge of interest in the development of automatic music generation systems (AMGSs; also known as algorithmic composition systems). Like human music composition and improvisation, AMGSs generally aim to create harmonic, timbral, and rhythmic sequences in an organized, musically-coherent fashion. This area, which sits at the intersection of computing, music theory/composition, and computational creativity, is relatively nascent, however, compared to the computational creation of visual art. This work aims to not only chip away at this gap, but offer a new automatic music generation system -- AffectMachine-Classical -- that is capable of producing controllable \textit{affective} music.  
AffectMachine-Classical offers an effective and flexible means of conveying emotions in real-time, and the system has been developed to be embedded into biofeedback systems such as brain-computer-interfaces (see for example \cite{ehrlich2019closed}), making it a potentially powerful tool for therapeutic applications such as emotion and mood regulation in listeners, augmentation of physical activity during rehabilitation, as well as commercial use cases such as soundtrack design and providing silent videos with novel music free of copyright. 

\subsection{Related work}

A review of automatic music generation is out of the scope of this article (for a review and summary of the state-of-the-art, see \citealt{herremans2017functional, carnovalini2020computational, dash2023ai}), however we will briefly summarize the main approaches to automatic music generation.
Previous approaches to developing music generation systems largely fall into two categories: learning-based methods, and rule-based methods. While there has been a recent trend towards learning-based approaches, they present several challenges for affective music generation. First, ecological (or realistic) music pieces typically exhibit hierarchical structure as well as polyphony. For example, a melodic phrase may extend over multiple measures of music, and involve several different instruments or voices. Further, music typically has an overall form that allows for musical and stylistic coherence throughout the piece. As such, a generative model must be able to capture harmonic, rhythmic and temporal structure, as well as the interdependency between voices \citep{dong2018musegan}. 
Second, learning-based approaches require large music datasets with emotion labels for training, a resource that is still scarce in the community. Style transfer models have had success as an alternative to models capable of generating novel affective music from scratch – for example, \citet{ramirez2015musical} used machine learning models to apply appropriate expressive transformations on the timing and loudness of pre-composed input musical pieces based on desired arousal and valence, while \citet{hu2020make} used convolutional neural networks to extract stylistic features from therapeutic music pieces and incorporate them into user-selected songs. However, style transfer models and similar approaches are subject to several important limitations. First, although leveraging pre-composed music greatly simplifies the challenge of producing affective music, the music chosen is subject to copyright. Second, these approaches are typically unable to support real-time interactivity, which is essential in biofeedback systems or any other systems meant to compose music in real-time to mediate the user's affective states.

In comparison to learning-based approaches, rule-based approaches rely on hand-designed functions to map affective signals to musical parameters. As such, they are able to sidestep the challenges associated with learning-based approaches by building in knowledge of how affective states map to musical parameters, as well as typical expectations regarding harmonic, rhythmic and temporal structure.
Additionally, the design of rule-based affective music generation systems benefits from an extensive body of theoretical and empirical work going back almost a century that investigates how different aspects of musical structure contribute to emotional expression \citep{gabrielsson2010role}. For example, the system described in \citet{wallis2011rule} was primarily informed by \cite{gabrielsson2001influence}, and generates novel music algorithmically by mapping seven musical parameters (e.g., note density, harmonic mode) to either valence or arousal in the most continuous possible way. Even though several salient musical parameters such as tempo, voice leading and voice spacing were not mapped for simplicity, the system was sufficient for participants to hear corresponding changes in the emotion of the music when changes were applied to the settings of valence and arousal parameters. Most recently, \cite{ehrlich2019closed} developed a system that loops over a I-IV-V-I harmonic progression, and modifies the harmonic mode, tempo, rhythmic roughness, overall pitch, and relative loudness of subsequent notes based on the desired level of valence and arousal.

\subsection{Emotion perception in music}
Music listening is often a rich emotional and cognitive experience \citep{altenmuller2012music}. Numerous empirical studies have been carried out to understand both the emotions that can be expressed through music (e.g., \cite{gabrielsson2003emotional}), as well as the musical factors that contribute to perceived emotional expression (e.g., \cite{gabrielsson2010role}). The relationship between musical factors and emotional expression is both relatively intuitive and well-understood. Tempo and mode, and to a lesser extent factors including musical dynamics, pitch range, loudness, rhythm and articulation, all influence the perceived emotion of a musical piece. Regarding the former, studies consistently find that listeners tend to exhibit agreement in their judgment of the general emotions expressed by a piece of music, and that these judgments are only marginally affected by demographic factors such as musical training, age and gender (e.g., \cite{juslin2004expression}, although see also  \cite{koh2023merp}). A notable feature of music is that it is often unable to reliably communicate finely differentiated emotions \citep{juslin1997can}. \citet{sloboda2000musical} offers an explanation for this phenomenon – music is to a large extent abstract and ambiguous, and while it may be able to suggest varying levels of energy or resemble certain gestures and actions, these emotional contours are often fleshed out in a subjective manner. 

Taken together, the literature suggests that (i) to a large extent, music can be systematically modified to express desired emotions, and that (ii) the effectiveness of affective music generation systems should be fairly robust across listeners.

\subsection{AffectMachine-Classical}
The current music generation system, AffectMachine-Classical, uses a probabilistic, rule-based approach to generate affective (classical) music in real time. 
Approaches to measuring the affect of musical stimuli have been fairly heterogenous, ranging from widely used measures such as \citet{russell1980circumplex}'s circumplex model and the Geneva Emotional Music Scale (GEMS; \citet{zentner2008emotions}) to bespoke methods developed for specific studies (e.g., \citet{costa2000psychological, lindstrom2006impact}).
Following much of the existing work on affective music systems (e.g., \cite{wallis2011rule, ehrlich2019closed}), we opt to represent emotion in AffectMachine using the circumplex model, in which emotions can be understood as points within a two-dimensional space. The first dimension is arousal, which captures the intensity, energy,  or “activation” of the emotion, while the second is valence, which captures the degree of pleasantness. For example, excitement is associated with high arousal and high valence, while contentment would be associated with low arousal and high valence. The circumplex model has several advantages over alternative measures of emotion. Firstly, to provide accurate and fine-grained feedback to a user about his or her emotional state, music generated by AffectMachine should ideally vary smoothly over the entire space of emotions, making continuous models of emotion such as the circumplex model a natural choice over categorical models of emotion such as GEMS. Secondly, allowing musical features to change gradually over time could help lend the music a more natural sound. Finally, the generalizability of the circumplex model also enables us to make use of previous research which may have used less common measures of emotion, by interpreting their results in terms of arousal and valence.

AffectMachine provides a model that is able to fluidly generate affective music in real time, either based on manually-input or predetermined arousal and valence values (e.g., as a sort of affective playlist for emotion mediation, or trajectory through emotion space), or based on the real-time feedback or physiological state of the user (e.g., EEG activity captured from the user and mapped to arousal and valence). In this way, AffectMachine offers a flexible yet powerful way to sonify (real-time) emotion states, and to influence the emotion states of the listener. 
The system may be used for health and wellness applications, such as generating affective playlists for emotion mediation. Further, AffectMachine may also be integrated into Brain-Computer Interface (BCIs) devices, or other systems capable of providing biofeedback, to assist the user in achieving a desired emotion state through neuro/biofeedback and affective music listening.

The main contributions of this work are: (1) the design of a novel rule-based affective music generation system to compose non-monotonic classical music, and (2) validation of the proposed system for expressing different emotions through a listener study. 
In the next section of this paper, we describe the features of AffectMachine-Classical (Section \ref{AMGS}). We then describe the listener study and discuss the findings and implications of our results (Section \ref{Study}), before providing our general conclusions and suggested future directions (Section \ref{GenDiscussion}).

\section{AffectMachine-Classical system description} \label{AMGS}
In this section, we describe the parameters and design of our novel affective music generation system, AffectMachine-Classical, which produces affective music in a classical style. AffectMachine was developed to be embedded in a BCI or neurofeedback system, to both generate emotion-inducing music in real-time, and to allow for neural or physiological signals (such as EEG) to \textit{drive} the music generation system. That is, the system was developed to both induce emotion in listeners, and provide users with real-time feedback on their current emotional state, in which the generated music is a reflection (or sonification) of the listeners' emotion state (when AffectMachine is embedded in a BCI or neurofeedback system). In the present paper, we remove AffectMachine from any embedded, interactive contexts (e.g., BCI), and examine the standalone AffectMachine, focusing on the efficacy of AffectMachine for generating music that conveys the intended emotion.

The automatic music generation system was developed in Python, and takes a sequence of arousal and valence states as input and encodes a corresponding sequence of harmonic, rhythmic and timbral parameters in the form of a MIDI event stream as output. The MIDI event stream is then sent to a digital audio workstation (DAW) over virtual MIDI buses to be translated into sound. For the present version of AffectMachine, we use the Ableton DAW for its wide selection of instruments and its ability to support live multi-track recording. Arousal and valence are continuous values within the range [0, 1] that can either be sampled from sensors (such as EEG) or manually provided. All musical parameters are updated each bar in accordance with the current arousal and valence values. 

Developing a rule-based affective music generation system requires first identifying a set of musical parameters and affective states, then designing functions that map parameter values to target states. For this reason, the harmonic, rhythmic, and timbral parameters were selected based on previous work establishing their influence on musical expression of emotions, and developed in collaboration with conservatory students formally trained in music composition.


In the subsections below, we present the details of the AffectMachine-Classical system.

\subsection{Harmonic parameters}

\subsubsection{Mode}

Previous rule-based music generation systems have controlled the mode parameter by choosing a fixed harmonic progression (e.g., I-IV-V-I) and in a few cases, by varying the harmonic mode from which the chords are drawn (e.g., each harmonic mode was mapped to a certain level of valence, with Lydian typically identified as the mode that expresses the highest valence and Locrian or Phrygian as the mode that expresses the lowest valence - a simpler, and much more common, version of this logic is to switch between the major and minor modes.)

In the AMG system, we introduce a completely novel way of controlling mode by using a bespoke probabilistic chord progression matrix inspired by the theme and variation form found in (human-composed) classical music. The music loops through an 8-bar theme with fixed chord functions for each bar, but the specific chords used, as well as their probabilities, are determined by the target level of valence desired. This approach is unobvious, and to our knowledge, has never before been implemented in a computational music generation system. Unlike previous systems which are constrained to a specific harmonic progression, the AMG system is extremely flexible – the only constraint being that the music has to progress through the 8-bar theme. (Note however that the majority of human-composed music also adheres to a repeating X-bar structure.) This novel approach is therefore beneficial by allowing a greater range of musical possibilities (and “interestingness” of the composition). At the same time, the music is able to achieve greater coherence of musical structure than what is commonly found in machine learning-based approaches by using chord substitutions in an 8-bar theme to express the desired level of valence, and by ending each iteration of the theme with a cadence.

To craft the 8-bar theme, the valence range was divided into ten regions, with one probabilistic chord progression composed for each region to match the intended level of valence. For example, at higher levels of valence, the chord progressions are composed in the major mode as it is typically associated with expressions of positive valence. As valence decreases, the likelihood of chords with greater tension or dissonance (such as those with diminished or minor intervals) increases. For a given bar (e.g., 1-8) and level of valence (e.g., 0-1), there are a set of possible chords, each with a particular probability of occurrence from 0.1 to 0.8. At any given bar and valence level, there are typically multiple chords (between one and five) to choose from.

\subsection{Pitch characteristics of voices}

\subsubsection{Voice leading}
Voice leading refers to the art of creating perceptually independent musical lines (e.g., tenor line, soprano line, etc) that combine to form a coherent piece \citep{huron2001tone}, and is a steadfast component of almost all human- composed polyphonic music. Despite the importance of voice-leading, automatic generation of polyphonic music with multiple voices or tracks is a challenge that research is only just beginning to address, primarily with learning-based generative methods (e.g., \cite{dong2018musegan}), and many of these systems either fail to address voice leading altogether or use highly simplified versions of voice leading.

In the AffectMachine system, we implement a novel rule-based music generation system that draws on both traditional rules of voice leading as well as heuristics used by human musicians, to create pieces that exhibit perceptually independent musical lines with nontrivial complexity and variability. By mapping these rules to differing levels of arousal and valence, we also provide more cues for listeners to identify the emotion being conveyed by the music, and enable finer-grained control over the mapping between affective states and musical parameters. This is an extremely important aspect and benefit of our approach. 

AffectMachine-Classical was developed to generate music with four parts or voices. The bass voice is carried by the string section and plays the root note of the current chord. The principal melody is placed in the soprano voice, which is carried by the clarinet and marimba. Both inner voices are carried by the piano, with the tenor voice simply playing the harmonic progression and the alto voice providing harmonic accompaniment. Instrumentation is explained in more detail in the section on timbral parameters.


While there are numerous principles that govern voice leading, or the creation of perceptually independent parts \citep{huron2001tone}, we select several straightforward rules that provide sufficient melodic diversity while minimizing unpleasant or artificial-sounding melodic lines. The three parts that are determined through voice leading logic are the tenor, alto and soprano voices. For the principal melody, our primary goal was to avoid unexpected dissonance. Hence, the note sequence is a randomly selected sequence of chord tones. For the tenor voice, which plays the harmonic progression, we follow the heuristic outlined in \citep{wallis2011rule} – that pianists tend to voice new chords in a manner that is as similar as possible to the previous chord, in terms of interval and placement on the keyboard. We calculate dissimilarity between two notesets ($N$, $N'$) as per Equation \ref{eqn:dissimilarity} and select the least dissimilar chord voicing to be played the first inner voice.

\begin{equation}\label{eqn:dissimilarity}
    dissimiliarity = \sum_i \sum_j \mid N_i - N'_j \mid \forall i \in N, \forall j \in N'
\end{equation}

For the alto voice, we combine chord tones with the step motion rule, which states that if the next note in the melody is of a different pitch, the pitch motion should be by diatonic step. These rules are encoded in the form of transition matrices. There are four possible states or motives: -1, indicating a diatonic step down the scale; 1, indicating a diatonic step up the scale; 0, indicating no pitch motion, and CT, indicating a jump to a randomly selected chord tone (CT). The arousal range was divided into two equal regions, with one matrix composed for each region to generate appropriate melodies for each level of arousal. The transition matrices were developed such that at higher levels of arousal, melodies are more likely to consist of scale patterns, mitigating the risk of the music being too dissonant or unpleasant due to the increased tempo and note density.

\subsubsection{Pitch register}
Research in the psychology of music has associated pitch height and pitch register with emotional expression for almost a century \citep{hevner1937affective}; yet pitch height is often not explicitly incorporated into automatic music generation systems.
Higher pitches generally tend to be associated with positively-valenced emotions such as excitement and serenity \citep{collier1998judgments}, while lower pitches tend to be associated with negatively-valenced emotions such as sadness. 

In AffectMachine-Classical, the pitch register of the lowest voice is consistent (at C3). For the remaining voices, the pitch register can vary within a permissible range determined by the current valence level.

To implement changes in pitch register, we again divided the valence range into ten equally spaced regions and tuned the lower and upper bounds of allowable pitches by ear. Both the lower and upper bounds of the range of permissible pitches increase gradually as valence increases. The range of permissible pitches starts at [C1, C5] in the lowest valence region, and gradually moves to [G3, C6] in the highest valence region.


\subsection{Time and rhythm parameters}

\subsubsection{Rhythm}
In most automatic generation approaches, the rhythmic content of the music is either fixed (e.g., a repeating pattern or a pre-composed rhythm template is used), or the temporal duration of notes (the rhythmic content) is based on a machine-learning generative process that affords little musical cohesion. This tends to either make the music sound extremely repetitive, or rather incoherent and unpleasant for most listeners.

To surmount this issue, the different voices/parts/tracks in AffectMachine-Classical use different rhythmic logic, e.g., one voice uses probabilistic rhythms while another uses composed rhythms. In this way, our new approach finds a nice and aesthetically-pleasant balance between composed and probabilistic elements.

As mentioned above, AffectMachine-Classical was developed to generate music using four parts or voices. The bass voice (string section) and first tenor voice (piano) employ a fixed rhythmic pattern -- they are both played on the first beat of each bar. For the soprano voice (clarinet and marimba), we divided the arousal range into three regions: low (Arousal $<$ 0.4), moderate (0.4 $\geq$ Arousal $<$ 0.75) and high (Arousal $>$ 0.75). Much like the implementation of mode, for a given bar (e.g., 1-8) and arousal region, there is a set of two possible rhythmic patterns or "licks" with equal probability of occurrence. The rhythmic pattern is represented in code as a list of binary values indicating whether each beat (subdivision) is associated with a note activation.


Finally, for the alto voice (piano), we incorporate rhythmic roughness, which is a measure of how irregular the rhythm of a piece of music is. Music with smooth, regular rhythms are typically perceived as higher in valence. In AffectMachine-Classical, we use note density as a proxy for rhythmic roughness \citep{wallis2011rule}. As arousal increases, roughness decreases and note density increases. When roughness is 0, each bar is populated with eight notes of equal length. However, this often results in overly dense-sounding output, because tempo is also high at higher levels of arousal. Hence, we limit the lowest roughness to 0.3.

\subsubsection{Tempo}
Tempo, or beats per minute, determines how quickly the notes of each bar are played. Alternatively, tempo can be thought of as a measure of note duration -- the faster the tempo, the shorter the note duration. In AffectMachine-Classical, tempo is determined by a simple linear relationship with arousal, and ranges from 60 bpm at Arousal = 0 to 200 bpm at Arousal = 1.

\subsection{Timbral and loudness parameters}
Two parameters contributed to variations in timbre: (i) the instrumentation of AffectMachine-Classical, and (ii) the velocity of notes, which refers to the force with which a note is played.

\subsubsection{Velocity range}
Similar to the algorithmic composition system developed by \citet{williams2017affective}, we mapped coordinates with higher arousal to brighter and harder timbres that were created by increasing MIDI key velocity. In MIDI, velocity is measured on a scale from 0-127. 
In our system, the range of permissible MIDI key velocities is [40, 70] at Arousal = 0, and the lower and upper bounds of the range increase linearly with arousal to [85, 115] at Arousal = 1. A uniform distribution over the range is used to determine the velocity for each bar.

\begin{equation}
Velocity = unif{40 + aro*45, 70+aro*45}
\end{equation}

\subsubsection{Velocity variation}
Patterns of velocity variation have affective consequences. For example, large variations are associated with fear, while small changes can indicate pleasantness. 

In our experimentation with the system, we found that frequent changes in velocity resulted in artificial, disjointed-sounding output. To strike a balance between having sufficient variation in velocity and incorporating those variations in as natural a way as possible, we limited the maximum change in velocity allowable within each bar. 

\subsubsection{Instrumentation}
Four virtual instruments were employed in the system (piano, a string section, clarinet, and marimba), and used to convey a classical musical style. As mentioned previously, the lowest voice is conveyed by the string section, while both inner voices are carried by the piano. The principal melody is placed in the uppermost voice, which is played by the clarinet. The marimba is used to double over the clarinet at high levels of valence (Valence $>$ 0.8) due to its cheerful-sounding timbre (and because, during experimentation with the system, marimba was found to nicely complement the timbre of the clarinet, which could sound slightly shrill at higher pitch heights). After all other harmonic, rhythmic and timbral parameters have been determined, instrument samples in the DAW (Ableton) are used to generate the final output audio.

\section{AffectMachine-Classical Listener study} \label{Study}

\subsection{Method}

A listening study was conducted in order to validate the efficacy of AffectMachine-Classical for generating affective music. We first used our system to generate brief musical examples from different points around the arousal-valence space of the circumplex model \cite{russell1980circumplex}. Listeners then provided arousal and valence ratings for each of these excerpts to examine whether the target emotion (in terms of arousal and valence) was indeed perceived as intended by listeners. 

\subsubsection{Participants}

The listening study was conducted with 26 healthy participants (average age = 22 yrs, SD = 4 yrs) including 11 male and 15 female participants. Twelve of the 26 participants reported having prior musical training. All the participants were given verbal and written instructions about the listening study prior to providing their written consent. The study was approved by the Institutional Review Board (IRB) of the National University of Singapore (NUS).


\subsubsection{Stimuli}

AffectMachine-Classical was designed to compose affective music that can span the entire valence-arousal plane. For the validation study, musical stimuli were generated from 13 different points around the valence and arousal plane. These were meant to represent different emotional states around the space, and covered the corners, middle of each quadrant, and the neutral middle point of the space. The points are: \{valence, arousal\} = [\{0,0\}; \{0,0.5\}; \{0,1\}; \{0.25;0.25\}; \{0.25,0.75\}; \{0.5,0\}; \{0.5,0.5\}; \{0.5,1\}; \{0.75,0.25\}; \{0.75,0.75\}; \{1,0\}; \{1,0.5\}; \{1,1\}]. There is a precedent in the literature for selecting these points in the arousal-valence plane for the validation of a music generation system \citep{ehrlich2019closed}.

To account for the probabilistic nature of the system, three different musical stimuli were generated from each of the thirteen points, resulting in a total of 39 musical excerpts. This mitigates the risk that artifacts in any particular stimulus might bias listener ratings, for more robust results. The average duration of the music stimuli is 23.6 seconds. The stimuli were composed based on either an 8-bar or 16-bar progression to allow the music to reach a cadence. While generating stimuli with a fixed duration is possible, this tends to result in stimuli that end very abruptly, which might influence a listener's emotional response to the stimuli. 16 bars were used for stimuli with a fast tempo (e.g., high arousal excerpts), as 8 bars produced too brief a time duration for these excerpts. All musical stimuli were presented to each participant in randomized order to avoid order effects across participants. The music stimuli used in this validation study are available online at \url{https://katagres.com/AffectMachineClassical_stimuli}.

\subsubsection{Experimental Protocol}

The experiment was conducted one participant at a time in a quiet room with minimal auditory and visual distractions. The experimenter first provided verbal and written instructions about the experiment, and then the participant provided written, informed consent to participate in the study. During the listening study, the participant sat in front of a computer and listened to the music stimuli over headphones, with the sound level adjusted to a comfortable listening volume. 


Before the listening task, the participant was asked to complete a demographic questionnaire which included questions about his/her age, prior musical training, ethnicity, etc. Subsequently, the participant rated his/her current emotional state. 

The music listening study began with two practice trials, followed by the 39 experimental trials in randomized order. After listening to each stimulus, the participant was asked to indicate the \textit{perceived} emotion of the stimulus (that is, the emotions they felt that the music conveyed) on a visual 9-point scale known as the Self-Assessment Manikin (SAM) \cite{bradley1994measuring}. These ratings were collected for both arousal and valence. Briefly, valence refers to the degree of the pleasantness of the emotion, while arousal refers to the activation or energy level of the emotion. The SAM scale ranged from `very unpleasant' (1) to `extremely pleasant' (9) for valence, and from `calm' (1) to `excited' (9) for arousal. Participants were allowed to take as long as they required to make these ratings, but were only permitted to listen to each musical stimulus once. The total duration of the experiment was approximately 40 minutes, and participants were compensated with \$6 SGD (equivalent to \$4.50 USD) for their time.

\subsection{Results and Discussion} \label{Results}

In order to evaluate the efficacy of the music generation system, we analysed the user ratings collected during the music listening study. We aimed to investigate (1) whether the music generated by the system is able to express the desired level of valence and arousal to the listeners, and (2) whether perceived valence and arousal are dependent on the listeners' prior musical training/knowledge. In this regard, we present our results in two subsections: (1) arousal and valence ratings, and (2) the impact of prior musical training on emotion ratings. We do not consider demographic factors such as age and ethnicity for further analysis due to the limited sample size.

As is commonly found in listener studies of emotion in music, we observed that the average valence and arousal ratings varied across listeners. This variance is often attributed to individual differences in musical preferences and training, and the listeners' demographic and cultural profile \citep{koh2023merp}. In order to mitigate the differences across listeners, we normalized the perceptual ratings from each user (see Equations 3 and 4 below). Here, $Max_{Valence}$ refers to the maximum possible valence rating (i.e., 9), and $Min_{valence}$ refers to the minimum possible valence rating (i.e., 1). The same $Max$ and $Min$ values apply to Arousal. The normalized valence and normalized arousal ratings, ranging between 0 to 1, are used for further analysis. In the remainder of the article, the normalized valence and normalized arousal ratings will be referred to as valence and arousal ratings, respectively.

\begin{equation}\label{eqn:Normalized_Valence}
Normalized_{Valence} = \frac{Rated_{Valence}}{(Max_{Valence} – Min_{Valence})}
\end{equation}

\begin{equation}\label{eqn:Normalized_Arousal}
Normalized_{Arousal} = \frac{Rated_{Arousal}}{(Max_{Arousal} – Min_{Arousal})}
\end{equation}

\subsection{Arousal and valence ratings}

\begin{figure}[h!]
\begin{center}
\includegraphics[width=12cm]{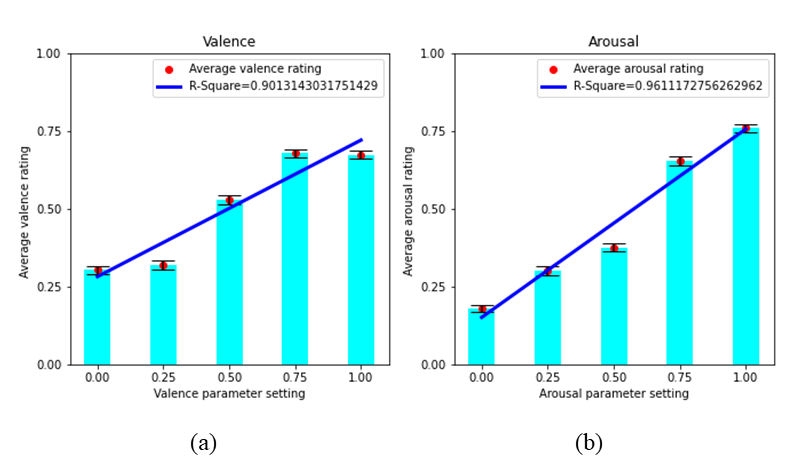}
\end{center}
\caption{(a) Linear regression between parameter settings and average valence ratings, (b) Linear regression between parameter settings and average arousal ratings. Error bars display standard error.}
\label{fig:1}
\end{figure}

To investigate whether AffectMachine is able to accurately express the intended emotion through music, we compared participants' averaged (normalized) emotion ratings for the musical stimuli with the valence or arousal parameter settings used during the music generation process. For example, the averaged valence ratings for all stimuli generated with the parameter settings \{valence, arousal\} = [\{0,0\}; \{0,0.5\}; \{0,1\}] were used to evaluate the system’s performance when valence is set to zero. The bar graphs depicting the averaged ratings (along with standard errors) are presented in Figure \ref{fig:1}. As expected, a strong increasing trend is seen for both the average valence and arousal ratings with respect to their corresponding parameter settings.
With regard to the valence ratings, we observe the majority of ratings to fall between the $<0.25$ and $>0.75$ parameter settings. It is common to see a higher density of responses in the middle of psychometric rating scales (e.g., with both ends of the scale receiving proportionally fewer responses) \citep{leung2011comparison}. This could also indicate that the extremes of the valence parameter values are less distinguishable by listeners. On the other hand, a better correspondence is observed between average arousal ratings and the respective parameter values at all levels of arousal.

To test the relationship between average valence and arousal user ratings and parameter settings, we performed linear regression analyses (illustrated in Figure \ref{fig:1}). The coefficient of determination is $R^2 = .90$ ($F =27, p<0.05$) for valence, and $R^2 = .96$ ($F=74,p<0.01$) for arousal, which confirms that both parameters are very effective in conveying their intended dimension of emotion. The results also show a stronger linear relationship for arousal (between average arousal ratings and parameter settings) in comparison to valence. This finding, in which arousal is more reliably expressed via music than valence, 
has previously been found in the literature \citep{ehrlich2019closed, wallis2011rule}. These results show that the music generated by AffectMachine-Classical generally conveys the intended levels of valence and arousal to listeners. 



\begin{figure}[h!]
\begin{center}
\includegraphics[width=15cm]{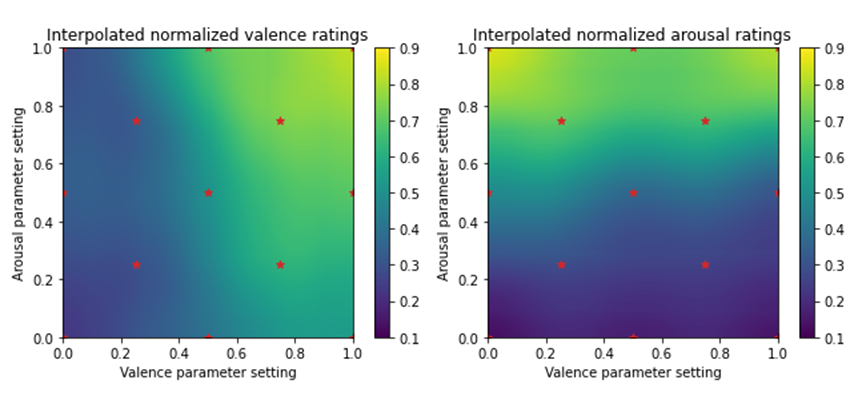}
\end{center}
\caption{Average (interpolated) valence and arousal ratings as a function of the valence and arousal parameters.}
\label{fig:interpolated}
\end{figure}

Next, we investigate whether the perception of valence is influenced by changes in the arousal parameter setting, and conversely whether the perception of arousal is influenced by changes in the valence parameter setting. To do so, we analyse the dependence of average emotion ratings on both the valence and arousal parameter settings together. Figure~\ref{fig:interpolated} visualizes this dependence by presenting the interpolated average valence (left) and arousal ratings (right) as a function of the emotion parameter settings. The stars in the figure represent the 13 points around the valence and arousal plane used to generate musical stimuli. As can be seen in the figure on the left, the perceived valence islower than the actual valence parameter setting (for $ V>0.7$) for excerpts expressing arousal values $<0.4$. That is, excerpts generated to express high valence convey only moderate valence when the 
arousal setting is low. This may be due in part to the effect of a slower tempo. Ratings at low valence settings are, however, in accordance with their respective parameter values. In contrast, we observe uniform correspondence between the arousal parameter values and arousal ratings regardless of the valence parameter setting. Our study replicates a phenomenon that has been previously described in \citet{wallis2011rule} -- the authors found asymmetrical ``crossover" effects between arousal and valence such that while perceived valence correlates with intended arousal, perceived arousal does not correlate significantly with intended valence.

To investigate these linear dependencies, we performed multiple linear regression between the valence and arousal parameter settings (independent variables) and average valence/arousal rating (dependent variable). The results indicate that perceived valence ratings are significantly influenced by both the valence ($ F=63, p < 0.001$) and arousal ($ F=11, p < 0.01$) parameter settings
, which is in line with what we observed in Figure~\ref{fig:interpolated}. Perceived arousal ratings, however, only show a significant dependence on the arousal settings ($ F= 153, p < 0.001$).
This observation is in line with findings from the literature which show that modelling the arousal component of emotion is more straightforward than the valence component \cite{wallis2011rule,yang2008regression}. Nevertheless, the obtained $R^2$ values are high $R^2 > 0.85$ for both average valence and arousal ratings. This confirms that irrespective of the emotion component, the majority of variability in average ratings during multiple regression analysis is explained by the valence and arousal settings values.

In summary, the listener study validates the ability of AffectMachine-Classical to generate music that expresses desired levels of emotion, measured in terms of arousal and valence. This confirms that the system has the potential to be deployed in applications that benefit from affective music - for example, the AffectMachine-Classical could be integrated with biofeedback systems wherein the music driven by the users' neural (or other physiological) signals can be used to reflect their emotional state. This direction is promising for developing more sophisticated emotion mediation systems with applications in healthcare \citep{agres2021music}. In the next section, we analyse the impact of participants’ prior musical training on emotion ratings. 



\subsection{Impact of prior musical training on emotion ratings}

In this section, we present a comparison of user ratings provided by participants with and without prior musical training. Participants indicated whether they had prior musical training in the demographic questionnaire they completed. Based on participants' response to the question ``Do you currently play an instrument (including voice)?” they were divided into two groups -- the musical training (MT) group and no musical training (NMT) group. The MT and NMT groups have 12 and 14 participants, respectively.

\begin{figure}[h!]
\begin{center}
\includegraphics[width=12cm]{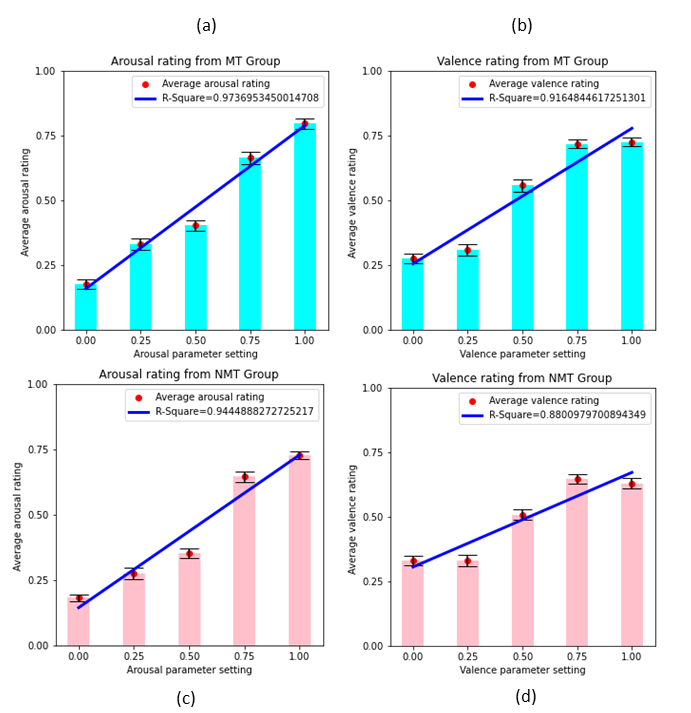}
\end{center}
\caption{(a) Average arousal rating and linear regression  for MT group, (b) Average valence rating and linear regression for MT group, (c) Average arousal rating and linear regression for NMT group, (d) Average valence rating and linear regression for NMT group. Error bars depict standard error.
}
\label{fig:5}
\end{figure}
Figure \ref{fig:5} presents the average emotion ratings corresponding to different levels of emotion parameter values for both the MT and NMT groups. As illustrated in the graphs, a stronger correspondence between the average emotion ratings and parameter-setting values is observed for arousal in comparison to valence, for both groups. As noted above, the average valence ratings demonstrate a saturation effect for lower ($<0.25$) and higher ($>0.75$) parameter-setting values for both the MT and NMT groups. Figure \ref{fig:5} also shows the linear regression fit for all the cases.
A comparison of the $R^2$ values reveals a stronger relationship between the emotion ratings and parameter settings for the MT group ($R^2$ for valence is $.91$, $F=33, p=0.01$; $R^2$ for arousal is $.97$, $F= 111, p<0.01$) than for the NMT ($R^2$ for valence is $.88$, $ F= 22, p<0.05$; $R^2$ for arousal is $.94$, $ F= 51, p<0.01$) group, for both valence and arousal. We may thus infer that prior musical training/knowledge enabled the MT group to more accurately recognise the target level of expressiveness in the affective music. This finding is in line with previous work, which highlights that musical expertise may influence the perception of emotion in affective music \citep{di2018effect, koh2023merp}. Regardless of musical training, however, all of the participants were able to reliably perceive the emotional expression in the music, which is evident from the high $R^2$ values observed ( ${>}$ 0.85) for both listener groups.

In addition, we also performed a multiple linear regression to examine the effect of parameter settings in both emotion dimensions on individual perceived emotion ratings. We obtained  high $R^2$ values ($R^2 > 0.8$) for all the scenarios, i.e., for both emotion dimensions for both groups. Furthermore, we observed that perceived valence is significantly influenced by both valence ($ F= 89, p<0.001$ for MT, 
and $ F= 38, p<0.001$ for NMT) and arousal ($ F=5.8, p<0.05$ for MT, and $ F=15, p<0.01$ for NMT) parameter settings for both MT and NMT groups. However, perceived arousal ratings are only influenced by the arousal settings ($ F=216, p<0.001$ for MT and $ F=110, p<0.001$ for NMT), and not valence settings, in both groups.
These findings are similar to what we observed for all the participants without any grouping.


\section{General Discussion} \label{GenDiscussion}


In this paper, we present a new computational system for generating affective classical music called AffectMachine-Classical. The system provides a probabilistic, rule-based algorithm for flexibly generating affective music in real-time. AffectMachine's behavior essentially resembles semi-structured musical improvisation, similar to how human jazz performers might follow the basic melody outlined by a lead sheet while coming up with reharmonizations, chord voicings and appropriate accompaniments, on the fly. To our knowledge, ours is the first affective music generation system to adopt this approach. A key advantage of this approach is that the music generated by the system exhibits strong coherence with minimal self-similarity, which may be valuable in research and other contexts that require lengthier pieces of music.
AffectMachine was developed to be embedded into a BCI system, an approach that leverages neurofeedback and affective music generation to help the listener achieve a target emotion state. The listener study reported here was conducted to validate the efficacy of the system for generating affective music.

The results of the listener study were promising, indicating a strong relationship between the arousal parameter setting and average arousal ratings ($R^2 = .96$), as well as the valence setting and average valence ratings ($R^2 = .90$). The correspondence between target and perceived emotion was more tempered for valence compared to arousal, as previously found in the literature (e.g., \cite{ehrlich2019closed, wallis2011rule}). 
From the results of our listener study, it is evident that the system is capable of expressing the desired emotional information and thus holds the potential to be used as an affect guide for mediating/transferring the mood states of the participants.
We would like to note here that despite the differences
in listeners’ prior musical training, individual and cultural preferences, and demographic profile, there was strong evidence suggesting that the system's target emotions were indeed perceived as intended by listeners, which makes Affect Machine-Classical a very promising tool for creating music with reliable emotion perception. 

In terms of future directions, as discussed above, AffectMachine will be embedded into biofeedback systems, such as a Brain-Computer-Interface (similar to \citep{ehrlich2019closed}, to support emotion self-regulation in listeners. Further, our system may be used for wellness applications such as generating affective music “playlist” for emotion mediation. That is, using the flexible music generation system, a user may pre-define an ‘emotion trajectory’ (e.g., a path through emotion space, such as the 2-dimensional Valence-Arousal space) to define the emotional qualities of their music over the duration of listening. For example, if a user desires 10 minutes of music to help him move from a depressed emotion state to a happy emotion state, he may indicate an emotion trajectory from negative arousal/valence to positive arousal/valence over the specified duration, and the system will create bespoke affective music to this specification. This is another prime example of how our system may be used for highly-personalized, affective music creation for well-being applications.

\section*{Conflict of Interest Statement}

The authors declare that the research was conducted in the absence of any commercial or financial relationships that could be construed as a potential conflict of interest.

\section*{Author Contributions}
KA and AD led the research. KA initiated and supervised the project. AD led data collection for the listener study and data analysis. PC led the system development, under the supervision of KA, and provided the system description. All authors contributed to writing the paper. 

\section*{Funding}
This work was supported by the RIE2020 Advanced Manufacturing and Engineering (AME) Programmatic Fund (No. A20G8b0102), Singapore.

\section*{Acknowledgments}
We would like to thank the composer Cliff Tan for his musical ideas, help in refining musical parameters, and assistance with DAW implementation when developing AffectMachine-Classical. We would also like to thank the composer Wen Liang Lim for contributing to our project ideation in the early stages of this work.




\bibliographystyle{Frontiers-Harvard} 
\bibliography{Refs}

\begin{thebibliography}{34}
\providecommand{\natexlab}[1]{#1}
\expandafter\ifx\csname urlstyle\endcsname\relax
  \providecommand{\doi}[1]{doi:\discretionary{}{}{}#1}\else
  \providecommand{\doi}{doi:\discretionary{}{}{}\begingroup
  \urlstyle{rm}\Url}\fi
\providecommand{\selectlanguage}[1]{\relax}
\providecommand{\bibAnnoteFile}[1]{%
  \IfFileExists{#1}{\begin{quotation}\noindent\textsc{Key:} #1\\
  \textsc{Annotation:}\ \input{#1}\end{quotation}}{}}
\providecommand{\bibAnnote}[2]{%
  \begin{quotation}\noindent\textsc{Key:} #1\\
  \textsc{Annotation:}\ #2\end{quotation}}

\bibitem[{Agres et~al.(2021)Agres, Schaefer, Volk, van Hooren, Holzapfel,
  Dalla~Bella et~al.}]{agres2021music}
Agres, K.~R., Schaefer, R.~S., Volk, A., van Hooren, S., Holzapfel, A.,
  Dalla~Bella, S., et~al. (2021).
\newblock Music, computing, and health: a roadmap for the current and future
  roles of music technology for health care and well-being.
\newblock \emph{Music \& Science} 4, 2059204321997709
\bibAnnoteFile{agres2021music}

\bibitem[{Altenm{\"u}ller and Schlaug(2012)}]{altenmuller2012music}
Altenm{\"u}ller, E. and Schlaug, G. (2012).
\newblock Music, brain, and health: exploring biological foundations of
  music’s health effects.
\newblock \emph{Music, health, and wellbeing} , 12--24
\bibAnnoteFile{altenmuller2012music}

\bibitem[{Bradley and Lang(1994)}]{bradley1994measuring}
Bradley, M.~M. and Lang, P.~J. (1994).
\newblock Measuring emotion: the self-assessment manikin and the semantic
  differential.
\newblock \emph{Journal of behavior therapy and experimental psychiatry} 25,
  49--59
\bibAnnoteFile{bradley1994measuring}

\bibitem[{Carnovalini and Rod{\`a}(2020)}]{carnovalini2020computational}
Carnovalini, F. and Rod{\`a}, A. (2020).
\newblock Computational creativity and music generation systems: An
  introduction to the state of the art.
\newblock \emph{Frontiers in Artificial Intelligence} 3, 14
\bibAnnoteFile{carnovalini2020computational}

\bibitem[{Collier and Hubbard(1998)}]{collier1998judgments}
Collier, W.~G. and Hubbard, T.~L. (1998).
\newblock Judgments of happiness, brightness, speed and tempo change of
  auditory stimuli varying in pitch and tempo.
\newblock \emph{Psychomusicology: A Journal of Research in Music Cognition} 17,
  36
\bibAnnoteFile{collier1998judgments}

\bibitem[{Costa et~al.(2000)Costa, Ricci~Bitti, and
  Bonfiglioli}]{costa2000psychological}
Costa, M., Ricci~Bitti, P.~E., and Bonfiglioli, L. (2000).
\newblock Psychological connotations of harmonic musical intervals.
\newblock \emph{Psychology of Music} 28, 4--22
\bibAnnoteFile{costa2000psychological}

\bibitem[{Dash and Agres(2023)}]{dash2023ai}
Dash, A. and Agres, K.~R. (2023).
\newblock Ai-based affective music generation systems: A review of methods, and
  challenges.
\newblock \emph{arXiv preprint arXiv:2301.06890}
\bibAnnoteFile{dash2023ai}

\bibitem[{Di~Mauro et~al.(2018)Di~Mauro, Toffalini, Grassi, and
  Petrini}]{di2018effect}
Di~Mauro, M., Toffalini, E., Grassi, M., and Petrini, K. (2018).
\newblock Effect of long-term music training on emotion perception from
  drumming improvisation.
\newblock \emph{Frontiers in Psychology} 9, 2168
\bibAnnoteFile{di2018effect}

\bibitem[{Dong et~al.(2018)Dong, Hsiao, Yang, and Yang}]{dong2018musegan}
Dong, H.-W., Hsiao, W.-Y., Yang, L.-C., and Yang, Y.-H. (2018).
\newblock Musegan: Multi-track sequential generative adversarial networks for
  symbolic music generation and accompaniment.
\newblock In \emph{Proceedings of the AAAI Conference on Artificial
  Intelligence}. vol.~32
\bibAnnoteFile{dong2018musegan}

\bibitem[{Ehrlich et~al.(2019)Ehrlich, Agres, Guan, and
  Cheng}]{ehrlich2019closed}
Ehrlich, S.~K., Agres, K.~R., Guan, C., and Cheng, G. (2019).
\newblock A closed-loop, music-based brain-computer interface for emotion
  mediation.
\newblock \emph{PloS one} 14, e0213516
\bibAnnoteFile{ehrlich2019closed}

\bibitem[{Fancourt and Finn(2019)}]{fancourt2019evidence}
Fancourt, D. and Finn, S. (2019).
\newblock \emph{What is the evidence on the role of the arts in improving
  health and well-being? A scoping review} (World Health Organization. Regional
  Office for Europe)
\bibAnnoteFile{fancourt2019evidence}

\bibitem[{Gabrielsson and Juslin(2003)}]{gabrielsson2003emotional}
Gabrielsson, A. and Juslin, P.~N. (2003).
\newblock \emph{Emotional expression in music.} (Oxford University Press)
\bibAnnoteFile{gabrielsson2003emotional}

\bibitem[{Gabrielsson and Lindstr{\"o}m(2001)}]{gabrielsson2001influence}
Gabrielsson, A. and Lindstr{\"o}m, E. (2001).
\newblock The influence of musical structure on emotional expression.
\bibAnnoteFile{gabrielsson2001influence}

\bibitem[{Gabrielsson and Lindstr{\"o}m(2010)}]{gabrielsson2010role}
Gabrielsson, A. and Lindstr{\"o}m, E. (2010).
\newblock The role of structure in the musical expression of emotions.
\newblock \emph{Handbook of music and emotion: Theory, research, applications}
  367400, 367--44
\bibAnnoteFile{gabrielsson2010role}

\bibitem[{Herremans et~al.(2017)Herremans, Chuan, and
  Chew}]{herremans2017functional}
Herremans, D., Chuan, C.-H., and Chew, E. (2017).
\newblock A functional taxonomy of music generation systems.
\newblock \emph{ACM Computing Surveys (CSUR)} 50, 1--30
\bibAnnoteFile{herremans2017functional}

\bibitem[{Hevner(1937)}]{hevner1937affective}
Hevner, K. (1937).
\newblock The affective value of pitch and tempo in music.
\newblock \emph{The American Journal of Psychology} 49, 621--630
\bibAnnoteFile{hevner1937affective}

\bibitem[{Hu et~al.(2020)Hu, Liu, Chen, Zhong, and Zhang}]{hu2020make}
Hu, Z., Liu, Y., Chen, G., Zhong, S.-h., and Zhang, A. (2020).
\newblock Make your favorite music curative: Music style transfer for anxiety
  reduction.
\newblock In \emph{Proceedings of the 28th ACM International Conference on
  Multimedia}. 1189--1197
\bibAnnoteFile{hu2020make}

\bibitem[{Huron(2001)}]{huron2001tone}
Huron, D. (2001).
\newblock Tone and voice: A derivation of the rules of voice-leading from
  perceptual principles.
\newblock \emph{Music Perception} 19, 1--64
\bibAnnoteFile{huron2001tone}

\bibitem[{Juslin(1997)}]{juslin1997can}
Juslin, P.~N. (1997).
\newblock Can results from studies of perceived expression in musical
  performances be generalized across response formats?
\newblock \emph{Psychomusicology: A Journal of Research in Music Cognition} 16,
  77
\bibAnnoteFile{juslin1997can}

\bibitem[{Juslin and Laukka(2004)}]{juslin2004expression}
Juslin, P.~N. and Laukka, P. (2004).
\newblock Expression, perception, and induction of musical emotions: A review
  and a questionnaire study of everyday listening.
\newblock \emph{Journal of new music research} 33, 217--238
\bibAnnoteFile{juslin2004expression}

\bibitem[{Koh et~al.(2023)Koh, Cheuk, Heung, Agres, and
  Herremans}]{koh2023merp}
Koh, E.~Y., Cheuk, K.~W., Heung, K.~Y., Agres, K.~R., and Herremans, D. (2023).
\newblock Merp: A music dataset with emotion ratings and raters’ profile
  information.
\newblock \emph{Sensors} 23, 382
\bibAnnoteFile{koh2023merp}

\bibitem[{Leung(2011)}]{leung2011comparison}
Leung, S.-O. (2011).
\newblock A comparison of psychometric properties and normality in 4-, 5-, 6-,
  and 11-point likert scales.
\newblock \emph{Journal of social service research} 37, 412--421
\bibAnnoteFile{leung2011comparison}

\bibitem[{Lindstr{\"o}m(2006)}]{lindstrom2006impact}
Lindstr{\"o}m, E. (2006).
\newblock Impact of melodic organization on perceived structure and emotional
  expression in music.
\newblock \emph{Musicae Scientiae} 10, 85--117
\bibAnnoteFile{lindstrom2006impact}

\bibitem[{Lonsdale and North(2011)}]{lonsdale2011we}
Lonsdale, A.~J. and North, A.~C. (2011).
\newblock Why do we listen to music? a uses and gratifications analysis.
\newblock \emph{British journal of psychology} 102, 108--134
\bibAnnoteFile{lonsdale2011we}

\bibitem[{MacDonald et~al.(2013)MacDonald, Kreutz, and
  Mitchell}]{macdonald2013music}
MacDonald, R., Kreutz, G., and Mitchell, L. (2013).
\newblock \emph{Music, health, and wellbeing} (Oxford University Press)
\bibAnnoteFile{macdonald2013music}

\bibitem[{Ramirez et~al.(2015)Ramirez, Palencia-Lefler, Giraldo, and
  Vamvakousis}]{ramirez2015musical}
Ramirez, R., Palencia-Lefler, M., Giraldo, S., and Vamvakousis, Z. (2015).
\newblock Musical neurofeedback for treating depression in elderly people.
\newblock \emph{Frontiers in neuroscience} , 354
\bibAnnoteFile{ramirez2015musical}

\bibitem[{Russell(1980)}]{russell1980circumplex}
Russell, J.~A. (1980).
\newblock A circumplex model of affect.
\newblock \emph{Journal of personality and social psychology} 39, 1161
\bibAnnoteFile{russell1980circumplex}

\bibitem[{Saarikallio(2012)}]{saarikallio2012development}
Saarikallio, S. (2012).
\newblock Development and validation of the brief music in mood regulation
  scale (b-mmr).
\newblock \emph{Music perception: an interdisciplinary journal} 30, 97--105
\bibAnnoteFile{saarikallio2012development}

\bibitem[{Sloboda(2000)}]{sloboda2000musical}
Sloboda, J.~A. (2000).
\newblock Musical performance and emotion: Issues and developments.
\newblock \emph{Music, mind, and science} , 220--238
\bibAnnoteFile{sloboda2000musical}

\bibitem[{Thayer et~al.(1994)Thayer, Newman, and McClain}]{thayer1994self}
Thayer, R.~E., Newman, J.~R., and McClain, T.~M. (1994).
\newblock Self-regulation of mood: strategies for changing a bad mood, raising
  energy, and reducing tension.
\newblock \emph{Journal of personality and social psychology} 67, 910
\bibAnnoteFile{thayer1994self}

\bibitem[{Wallis et~al.(2011)Wallis, Ingalls, Campana, and
  Goodman}]{wallis2011rule}
Wallis, I., Ingalls, T., Campana, E., and Goodman, J. (2011).
\newblock A rule-based generative music system controlled by desired valence
  and arousal.
\newblock In \emph{Proceedings of 8th international sound and music computing
  conference (SMC)}. 156--157
\bibAnnoteFile{wallis2011rule}

\bibitem[{Williams et~al.(2017)Williams, Kirke, Miranda, Daly, Hwang, Weaver
  et~al.}]{williams2017affective}
Williams, D., Kirke, A., Miranda, E., Daly, I., Hwang, F., Weaver, J., et~al.
  (2017).
\newblock Affective calibration of musical feature sets in an emotionally
  intelligent music composition system.
\newblock \emph{ACM Transactions on Applied Perception (TAP)} 14, 1--13
\bibAnnoteFile{williams2017affective}

\bibitem[{Yang et~al.(2008)Yang, Lin, Su, and Chen}]{yang2008regression}
Yang, Y.-H., Lin, Y.-C., Su, Y.-F., and Chen, H.~H. (2008).
\newblock A regression approach to music emotion recognition.
\newblock \emph{IEEE Transactions on audio, speech, and language processing}
  16, 448--457
\bibAnnoteFile{yang2008regression}

\bibitem[{Zentner et~al.(2008)Zentner, Grandjean, and
  Scherer}]{zentner2008emotions}
Zentner, M., Grandjean, D., and Scherer, K.~R. (2008).
\newblock Emotions evoked by the sound of music: characterization,
  classification, and measurement.
\newblock \emph{Emotion} 8, 494
\bibAnnoteFile{zentner2008emotions}

\end{thebibliography}

\end{document}